\documentclass{ws-ijmpe}

\usepackage{bm}% bold math
\usepackage{graphicx}

\newcommand{\be}{\begin{equation}}
\newcommand{\ee}{\end{equation}}
\newcommand{\bn}{\begin{eqnarray}}
\newcommand{\en}{\end{eqnarray}}
\newcommand{\ba}{\begin{array}}
\newcommand{\ea}{\end{array}}
\newcommand{\bc}{\begin{center}}
\newcommand{\ec}{\end{center}}
\newcommand{\bml}{\begin{mathletters}}
\newcommand{\eml}{\end{mathletters}}

\begin{document}

\markboth{W. Satu{\l}a }{Probing effective NN interaction at band
termination}

\catchline{}{}{}{}{}
%
%%%%%%%%%%%%%%%%%%%%%%%%%%%%%%%%%%%%%%%%%%%%%%%%%%%%%%%%%%%%%%%%%%%%

\author{W. SATU{\L}A\footnote{satula@fuw.edu.pl}}
\address{Institute of Theoretical Physics, University of Warsaw, \\
ul. Ho\.za 69, 00-681 Warsaw, Poland,\\
%KTH (Royal Institute of Technology)\\
%AlbaNova University Center, 106 91 Stockholm, Sweden\\
}

\title{PROBING EFFECTIVE NUCLEON-NUCLEON INTERACTION AT BAND TERMINATION}

\maketitle
\begin{history}
\received{(received date )}
\revised{(revised date )}
%\accepted{(Day Month Year)}
%\comby{(xxxxxxxxxx)}
\end{history}

\begin{abstract}
Low-energy nuclear structure is not sensitive enough to
resolve fine details of nucleon-nucleon (NN) interaction.
Insensitivity of infrared physics to the details of short-range strong
interaction allows for consistent, free of ultraviolet divergences,
formulation of local theory  at the level of local energy density
functional (LEDF) including, on the same footing, both particle-hole as
well as particle-particle channels. Major difficulty is related to
parameterization of the nuclear LEDF and its density dependence. It is argued
that structural simplicity of terminating or isomeric states offers invaluable
source of informations that can be used for fine-tuning of the NN interaction
in general and the nuclear LEDF parameters in particular. Practical
applications of terminating states
at the level of LEDF and nuclear shell-model are discussed.
\end{abstract}

\section{Introduction}

 The atomic nuclei are very complex finite many-body systems
 exhibiting non-trivial coupling of single-particle (sp)
 and collective degrees of freedom.
 It is well known that significant fraction of these
 many-body correlations can be taken into account through
 symmetry violating intrinsic states within self-consistent
 mean-field (MF) approximation using finite-range
 Gogny~\cite{[Gog73x]} or
 contact Skyrme~\cite{[Sky56x]} interactions.
 These interactions, or the underlying energy density
 functionals (EDF), are parametrized usually by about
 ten coupling constants which are adjusted to basic properties
 of nuclear matter and to selected data on finite nuclei.

 The fitting procedure is by no means unambiguous.
 There is no clear consensus of what dataset should be used
 in adjusting EDF parameters.  Coupling between
 single-particle and collective degrees of freedom and
 correlations beyond MF manifestly appearing in
 any finite mesoscopic system
 implies further that different philosophy (procedures and datasets)
 should, in principle, be applied while
 fitting EDF for pure MF applications and for
 theoretical methods taking explicitly correlations beyond
 MF approximation like random phase approximation, generator-coordinate
 method, or symmetry-projection techniques to avoid double counting.
 Yet another open problem is related to density
 dependence of nuclear EDF coupling constants and its relation to
 the NNN interaction.  Presently used prescription
 is purely phenomenological. It is not only
 unsatisfactory from theoretical point of view but
 cause numerous troubles in particular in angular momentum
 projected calculations, see~\cite{[Kaz06]} and refs. quoted therein.
 Similar ambiguities related to the form and density dependence
 apply to pairing channel.

 The number of existing parameterizations of, in particular, Skyrme
 forces reflects ambiguities listed above. This rather
 frustrating situation can be healed to a certain extent by
 turning the attention to specific, extremely simple nuclear
 states where pairing, single-particle and collective degrees
 of freedom decouple to the largest possible extent. The classical examples
 of such states are superdeformed states or terminating and high-spin
 isomeric states, see~\cite{[Afa99a],[Sat05]} and refs. cited therein.

 \smallskip

Hereafter I will focus on applications of terminating and high-spin
isomeric states. The aim is to demonstrate how the
structural simplicity of these states
can be used to  probe nuclear Skyrme EDF
(see Ref.~\cite{[Zdu05y],[Zal07]} for further details),
pairing correlations as well as
effective $sdfp$ shell-model (SM) NN interaction.

I will start by presenting general arguments
speaking in favor of local nuclear theory. I will show next a couple of
numerical results pertaining to isovector pairing correlations.
In particular, I will consider three versions of density dependent
delta interaction (DDDI) including volume-active, surface-active
and mixed variants. I will present examples
of calculated shape-gap correlation plots that seem to quite
firmly point out toward volume-like character of
nucleonic isovector pair field by excluding both
surface-active or mixed pairing scenarios.
Next, I will briefly overview problems with conventional
pair-blocking encountered in the analysis of
$N$=83 high-spin isomeric states in rare-earth
nuclei. Finally, I will discuss the energy differences
between two terminating configurations
$[f_{7/2}^n]_{I_{max}}$ and
$[d_{3/2}^{-1} f_{7/2}^{n+1}]_{I_{max}}$ in
 $Z$$\leq$$N$,  $A$$\sim$45 mass region
\begin{equation}\label{deltae}
\Delta E = E([d_{3/2}^{-1} f_{7/2}^{n+1}]_{I_{max}}) -
E([f_{7/2}^n]_{I_{max}}).
\end{equation}
This extremely simple observable
can be used not only to test time-odd spin-fields and spin-orbit
strength of the nuclear EDF~\cite{[Zdu05y],[Zal07]} but
also elucidates much more subtle effects directly pertaining
to isospin dependence of SM matrix elements.

\section{Infrared nuclear theory}

\subsection{Particle-hole channel}

Nuclear structure theory aims to describe low-energy nuclear
excitations. Since such observables are be definition not sensitive enough
to resolve details of the underlying NN (and NNN)
interaction one can freely relax any less or more stringent
relation between effective and
{\it ab initio\/} NN (and NNN) potentials and
attempt to built the effective nuclear theory
essentially from scratch using, as the only guiding principle,
the following general statement: {\it
low-energy or\/} INFRARED {\it physics should not depend on high-energy
or\/} ULTRAVIOLET {\it dynamics\/}. Such separation of scales
is the underlying  principle of effective field theory
which aims to describe composite objects
at low-energies by Largrangians which include ultraviolet dynamics
by means of a serie of contact corrections~\cite{[Lap97x]}.
It implies that the following
expansion of an arbitrary short-range, rotationally invariant
NN interaction:
 \be\label{vq}
 v_S(q^2) \approx v_S(0) + v_S^{(1)} (0) q^2  + v_S^{(2)} (0) q^4
 \ldots\, ,
 \ee
in terms of transferred momentum $q$ should
converge fast.\footnote{At this stage we omit,
for the sake of simplicity, spin and isospin degrees of freedom.} In this way
complicated and in fact unknown two-body NN interaction is mapped by just a
few constants: $v_S(0), v_S^{(1)} (0), \ldots$ which should be
carefully adjusted using representative set of nuclear data.

In the $r$-space the expansion (\ref{vq}) takes the following form
 \bn\label{vcorr}
    v_{eff}({\boldsymbol r}) &\approx & v_{long} ({\boldsymbol r}) \nonumber \\
 & + & c a^2 \delta_a ({\boldsymbol r}) \nonumber \\
 & + & d_1 a^4 {\boldsymbol \nabla}^2 \delta_a ({\boldsymbol r})
   +  d_2 a^4 {\boldsymbol \nabla} \delta_a ({\boldsymbol r})
   {\boldsymbol \nabla}\nonumber \\
 & + & \ldots \nonumber \\
 & + & g_1 a^{n+2} {\boldsymbol \nabla}^n \delta_a ({\boldsymbol r}) +\ldots
\, ,
\en
 where  $\delta_a ({\boldsymbol r})$ denotes
 an arbitrary model of the Dirac delta function of range $a$.
 The $v_{long} ({\boldsymbol r})$
 describes the long-range part of the NN potential.
 Since typical range of strong interaction is $a\sim 1$\,fm and is
 much smaller than nuclear radius $R$, $a\ll R$, the
 $v_{long} ({\boldsymbol r})$ pertains essentially
 to Coulomb interaction. The correcting terms are proportional to
 $\delta_a ({\boldsymbol r})$ and derivatives of
 $\delta_a ({\boldsymbol r})$ arranged as in Eq.~(\ref{vcorr})
 to assure spherical symmetry.  Each correcting
 term  introduces its own dimensionless coupling constant
 $c, d_1, d_2\ldots g_1\ldots$ since
 the effective range $\sim a$ or ultraviolet
 cut-off momentum $\sim 1/a$ is pulled out explicitly in
 Eq.~(\ref{vcorr}).
 As already mentioned these coupling constants
 should be readjusted to a selected set of low-energy data.
% Moreover they should obey a criterion of naturalness meaning that
% expansion should converge order-by-order.

 The assumed spherical symmetry
 defines  uniquely (Eq.~(\ref{vcorr})) the form of the correcting potential
 replacing the low-momentum part of the exact potential removed by
 the ultraviolet cut-off procedure. However, since
 $\delta_a ({\boldsymbol r})$ can be modeled in essentially
 arbitrary way, there is, at least in principle, an infinite set of
 equivalent realizations of the effective interactions or
 effective theories.  In particular, assuming Gaussian
form factor:
 \be\label{gauss}
   \delta_a ({\boldsymbol r})\equiv
   \frac{\mbox{e}^{-{\boldsymbol r}^2/2a^2}}{(2\pi)^{3/2}a^3 }\, ,
 \ee
 or more precisely modeling $\delta_a ({\boldsymbol r})$  by a sum
 of attractive and repulsive Gaussians
 of different ranges with space-, spin- and isospin-exchange term
 and  supplementing it by a density dependent term and a
 spin-orbit term (last two terms in Eq.~(\ref{gogny})) leads to the
 Gogny force~\cite{[Gog73x]}
 \bn\label{gogny}
 v({1,2}) & = & \sum_{j=1}^2
 e^{{\boldsymbol r}_{12}^2/\mu_j^2} \left( W_j -
 B_j \hat{P_\sigma}   - H_j \hat{P_\tau}  - M_j\hat{P_\sigma}\hat{P_\tau}
   \right) \nonumber \\
 &+&  t_3 (1+x_3 \hat P_\sigma) \rho_{0}^\gamma
  ({\boldsymbol R})
  %\left(\frac{{\boldsymbol r_1}+{\boldsymbol r_2}}{2}\right)
  \delta ({\boldsymbol r}_{12}) \nonumber \\
 &+& iW_0({\boldsymbol \sigma_1}+{\boldsymbol \sigma_2} )
  \left( \hat{\boldsymbol k}^\prime \times
  \delta ({\boldsymbol r}_{12}) \hat{\boldsymbol k} \right)\, .
 \en

\bigskip

Since the early work of Brink and Vautherin~\cite{[Vau72]} it is known
that particle-hole (ph) channel can be described by contact
$\lim_{a \rightarrow 0} \delta_a = \delta
({\boldsymbol r}) $ Skyrme interaction~\cite{[Sky56x]}:
 \bn\label{skyrme} v({1,2}) & = &
    t_0 (1+x_0 \hat P_\sigma) \delta ({\boldsymbol r}_{12}) \nonumber \\
 & +&  \frac{1}{2} t_1 (1+x_1 \hat P_\sigma) \left(
      \hat{\boldsymbol k}^{\prime 2}  \delta ({\boldsymbol r}_{12})
       + \delta ({\boldsymbol r}_{12}) \hat
       {\boldsymbol k}^{2} \right) \nonumber \\
 &+& t_2 (1+x_2 \hat P_\sigma) \hat{{\boldsymbol k}}^\prime
  \delta ({\boldsymbol r}_{12}) \hat{{\boldsymbol k}} \nonumber  \\
 &+& \frac{1}{6} t_3 (1+x_3 \hat P_\sigma) \rho_{0}^\gamma
  ({\boldsymbol R})
  \delta ({\boldsymbol r}_{12}) \nonumber \\
 &+& iW_0({\boldsymbol \sigma_1}+{\boldsymbol \sigma_2} )
  \left( \hat{\boldsymbol k}^\prime \times
  \delta ({\boldsymbol r}_{12}) \hat{\boldsymbol k} \right)\, ,
 \en
 where: ${\boldsymbol r}_{12}= {\boldsymbol r_1}-{\boldsymbol r_2}$;
 ${\boldsymbol R}= ({\boldsymbol r_1}+{\boldsymbol r_2})/2$;
 the momentum operator  $\hat{{\boldsymbol k}}= \frac{1}{2i} ({\boldsymbol
 \nabla}_1 -  {\boldsymbol \nabla}_2 )$ acts to the right while
 $\hat{{\boldsymbol k}}^\prime = - \frac{1}{2i} ({\boldsymbol
 \nabla}_1 -  {\boldsymbol \nabla}_2 ) $ acts to the left hand side.
 %$\hat{P}_\sigma = \frac{1}{2}(1+{\boldsymbol \sigma_1}
 %{\boldsymbol \sigma_2})$ denotes the spin-exchange operator.
 The first three terms of the Skyrme interaction correspond to the first
 three terms in the expansion (\ref{vcorr}). Last two terms in
 Eq.~(\ref{skyrme}) denote density dependent and
 spin-orbit terms.

\bigskip

 The concept of effective theory based on renormalization of
 ultraviolet dynamics turns up side down the philosophy
 behind the effective NN forces.
 It states that due to {\it poor\/} resolution of low-energy data
 the exact form of the NN force is not at all needed in practical
 computations. It can be replaced by local corrections of the form
 given in Eq.~(\ref{vcorr}). The insensitivity of the effective
 theory to the short-range details tells us that we can construct
 infinitely many theories having the same low-energy behavior.
 All of them (although different and having probably not much
 in common  with the true short-range part of the NN interaction) are
 essentially equivalent in the sense that all of them should be
 capable to reproduce low-energy nuclear data with desired
 accuracy when order-by-order refinement~(\ref{vcorr}) is applied.
 In this sense Gogny and Skyrme
 interactions appear as two independent (among infinitely many)
 realizations of an effective theory.
 So far, Skyrme interaction was viewed rather as a short-range expansion
 of finite-range Gaussian force~\cite{[Vau72]} and regarded
 as a limiting case of a
 seemingly more fundamental finite-range Gogny force.

\subsection{Particle-particle channel}

\begin{figure}[t]
\par
\centerline{\epsfig{file=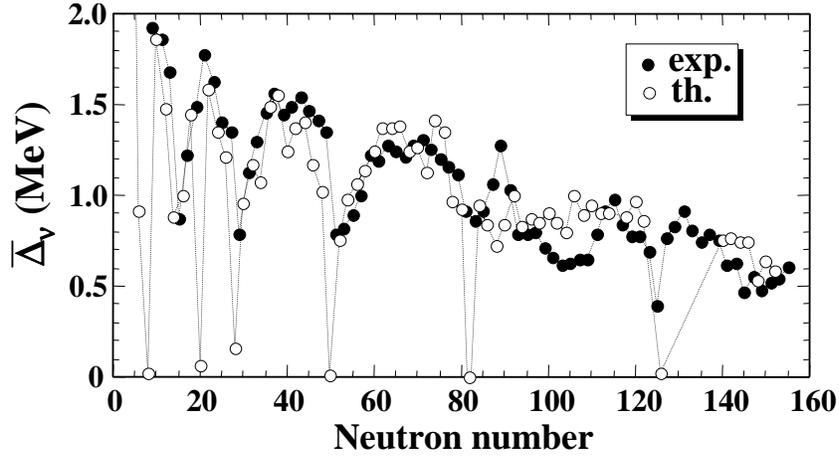,height=6cm,clip}}
\vspace*{8pt}
\caption{
Mean neutron pair gaps calculated using HFB D1S Gogny
method (open dots) in Ref.~\protect\cite{[Hil02]}
in comparison
to empirical three-point OES data $\Delta (odd-N)$ (filled dots).}
\label{gog}
\end{figure}

 The particle-hole and particle-particle (pp) or pairing interactions
can be treated independently. One can therefore apply similar
expansion to the pp interaction:
\be\label{vpair}
v_{pair}(q^2) \approx g + g_2 q^2 + g_4 q^4 \ldots
\ee
By retaining only the first term in (\ref{vpair})
and by modeling it by a Gaussian-type Dirac-delta model
(\ref{gauss}) one obtains finite range Gogny pairing force
which appeared to be very successful in numerous practical
applications. The example of such calculations illustrating
Hartree-Fock-Bogolyubov (HFB) mean neutron gap calculated
using D1S Gogny force versus neutron number~\cite{[Hil02]} is shown
in Fig.~\ref{gog}. Note, that the theoretical pair-gaps
follow very closely empirical three-point
odd-even staggering (OES) data:
\be\label{3point}
\Delta (N) = \frac{(-1)^N}{2} [ B(N-1)+B(N+1) - 2B(N) ] \, ,
\ee
extracted for odd-$N$ nuclei which, in accordance with the
analysis of Refs.~\cite{[Sat98],[Dob01]}, represents mean pair gap.
In particular no low-mass enhancement of the pair gap following
conventional textbook $\Delta \sim 1/\sqrt{A}$ estimate is seen neither in the
data nor in the calculations. An additional advantage of finite-range
pairing model is that it automatically removes high-momentum
scattering processes. Indeed, due to the finite-range $r_o$ ($r_o\sim 1$\,fm)
the Gogny force discriminates states above
$E_c\sim p_c^2/2m_r \sim \hbar^2/mr_o \sim 40$\,MeV [$m_r=m/2$ is
reduced mass] since $r_o p_c\sim \hbar$.

\bigskip

In spite of its success
the use of finite range interaction in the pp channel may be viewed
as rather unnecessary complication.
Indeed, in the nuclear matter pairing modifies
the nucleonic motion essentially only in the closest vicinity of the Fermi
energy, $E_F - \Delta \leq
\frac{(p_F\pm \delta p)^2}{2m} \leq E_F + \Delta$
giving rise to uncertainty in momentum space,
$\delta p \sim \Delta/v_F$.  This uncertainty translates to uncertainty
in coordinate space
$\xi \sim \frac{(\hbar c)^2 k_F}{(m c^2)\Delta} \gg r_o \sim
\frac{1}{k_F}$ which by far exceeds the typical interaction range $r_o$.
Consequently, the nucleonic Cooper
pairs appear as spatially extended objects which can hardly feel
subtle details of the underlying
NN pairing interaction. Nuclear pairing should be therefore well
described within local approximation using plain delta interaction or,
slightly more general, density dependent delta interaction (DDDI):
\be\label{dddi}
v_{pair} ({\boldsymbol r}) = v_o \left[ 1 -
\left( \frac{\rho ({\boldsymbol r})}{\rho_c} \right)^\alpha \right]
\delta ({\boldsymbol r}).
\ee

\bigskip

\begin{figure}[t]
\par
\centerline{\epsfig{file=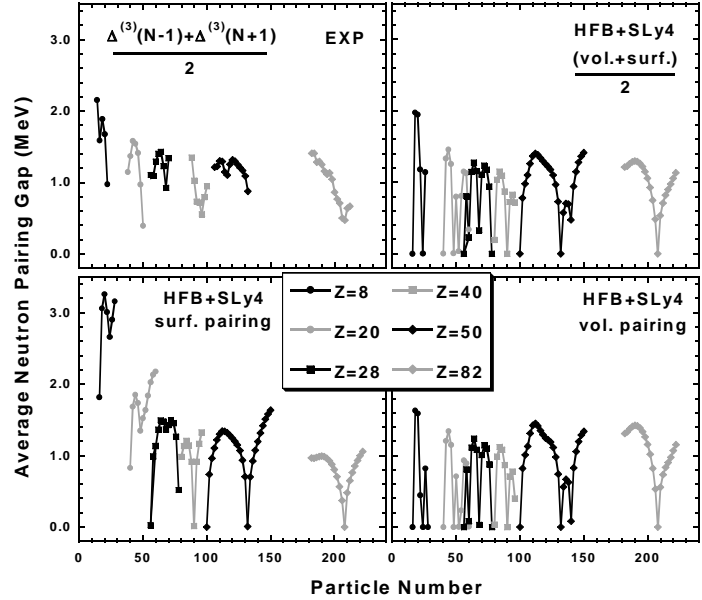,height=8cm,clip}}
\vspace*{8pt}
\caption{Neutron pair-gaps versus neutron number. Upper left panel
shows empirical OES. The remaining panels illustrate pair-gaps
obtained within Skyrme-HFB model using three different versions
of the DDDI interaction:
surface-active (lower left), mixed
(upper right) and volume-active (lower right).
Taken from Ref.~\protect\cite{[Dob02x]}.
}
\label{jacek}
\end{figure}

We are again touching a subtle question of resolution and sensitivity of low
energy data but this time with respect to pairing. In particular case of
nuclear matter it was demonstrated explicitly by Garrido {\it et
al.\/}~\cite{[Gar99]} that one can rather easily parametrize interaction
(\ref{dddi}) in order to reproduce pair-gap $\Delta (k_F)$
versus $k_F$ dependence in
nuclear matter (with free particle spectrum) obtained using Gogny interaction.
In the case of finite nuclei the situation is far more complicated.
Figure~\ref{jacek} illustrates empirical neutron-gap (OES) and the results of
large scale spherical Skyrme-HFB calculations using DDDI interaction in pp
channel~\cite{[Dob02x]}.
Three different versions of the calculations are
depicted corresponding to surface ($\rho_c = \rho_o$), mixed ($\rho_c =
2\rho_o$) and volume ($\rho_c \rightarrow \infty$) type pairing scenarios. In
all cases $\alpha =1$. The strength $v_o$ was adjusted separately for each
pairing-mode in order to reproduce empirical neutron pair-gap in
$^{120}$Sn.

The overall agreement between neutron OES and the calculated pair-gaps
is satisfactory for volume-active and mixed pairing scenarios.
For these two pairing scenarios the agreement is also
similar to the one obtained with Gogny-pairing shown in Fig.~\ref{gog}.
Clearly, the OES itself is not {\it sensitive\/} enough to differentiate
neither between these two variants of the DDDI interaction nor
between the DDDI and Gogny pairing.
On the other hand the calculations clearly prefer volume or mixed
DDDI over pure surface-active DDDI. Indeed, the
surface-active pairing leads to enhanced neutron gaps
in light systems which are not at all observed in the data.

\begin{figure}[t]
\par
\centerline{\epsfig{file=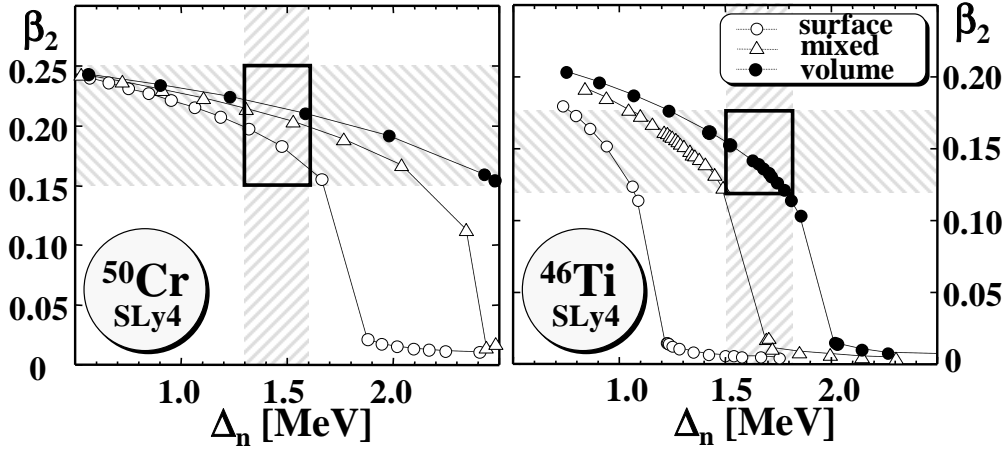,height=6cm,clip}}
\vspace*{8pt}
\caption{Shape-gap correlation plot in $^{50}$Cr (left) and
$^{46}$Ti (right). Three different curves represent Skyrme-HFB
calculations assuming surface-active (open dots), mixed
(triangles) and volume-active (black dots) pairing variants.
}
\label{kosm}
\end{figure}

Further inside into a character of nuclear pairing can be gained
by looking into correlation plots of deformation, which is rather
robust observable, versus pairing gap. An example of such
shape-gap consistency plot is shown in Fig.~\ref{kosm}.
Two different examples of rather well deformed nuclei are
depicted in the figure. Left part shows $\beta_2$-vs-$\Delta_n$ plot
in $^{50}$Cr. This is a typical case when shape-gap consistency
between calculations and the data
(marked by the area where horizontal and vertical
shaded areas corresponding to empirical uncertainties
in $\beta_2$ and $\Delta_n$ cross each other)
can be reached essentially irrelevant of the
assumed pairing scenario. Hence, this example shows {\it no sensitivity\/}
at all with respect to the considered pairing variant.
The right hand side of Fig.~\ref{kosm} shows similar
$\beta_2$-vs-$\Delta_n$ plot but for $^{46}$Ti.
In this case
shape-gap consistency requirement simply {\it excludes\/} both
the surface-active and mixed pairing scenarios.  There
are at least two almost obvious questions which need to be answered
in this context: ({\it i\/}) Can the hierarchy of deformed-to-spherical
phase transition versus pairing type which so
clearly seen in Fig.~\ref{kosm} be understood in a simple manner?
({\it ii\/}) Is the case exclusiveness in $^{46}$Ti
accidental or it can be traced down systematically throughout the periodic
table? Studies along these lines are under way.

\subsection{Toward consistent superfluid local density approximation}

 The disadvantage of using DDDI type (local) pairing interaction as compared
to finite-range Gogny interaction is that the former requires explicit
cutoff parameters to avoid divergences. In particular, all
Skyrme-HFB calculations presented above were done within limited
phase-space for pairing. Although practically sufficient
in most applications, such a brute-force method is
 unsatisfactory from theoretical point of view.
In this respect, local pairing can be considered as considerably
disadvantageous as compared to finite-range pairing. It appears however,
that the divergence can be rather easily identified and subsequently
regularized leading to consistent cutoff-free superfluid local density
approximation (SLDA)~\cite{[Pap99],[Bul02x]}.

The appropriate scheme for regularization of ultraviolet divergence
in anomalous density matrix:
 \be
   \nu ({\boldsymbol r_1}, {\boldsymbol r_2}) =
   \sum_i v_i^*({\boldsymbol r_1}) u_i ({\boldsymbol r_2})
   \sim  \frac{1}{|{\boldsymbol r_1} - {\boldsymbol r_2} |}\, ,
\ee
at the level of  the local density approximation (LDA)
i.e. including properly dominant particle-hole channel
was recently proposed by Bulgac and Yu~\cite{[Bul02x]}.
The idea is to renormalize divergent terms by
introducing cutoff ($E_c \equiv \frac{(\hbar k_c)^2}{2m}$) dependent
counter-terms. Such an approach leads
to standard local HFB formalism with cutoff
parameters but with a gap equation dependent on the effective
{\it running\/} coupling constant:
\bn
   \nu_c ({\boldsymbol r}) & = & \sum_{E_i \geq 0}^{E_c}
    v_i^*({\boldsymbol r}) u_i ({\boldsymbol r})\, ,    \\
   \Delta ({\boldsymbol r}) & = &  -g_{eff} ({\boldsymbol r})
   \nu_c ({\boldsymbol r}) \label{delta2}\, , \\
   \frac{1}{g_{eff} ({\boldsymbol r})} & = &
   \frac{1}{g[\rho ({\boldsymbol r})]} - \frac{m ({\boldsymbol r})
   k_c ({\boldsymbol r})}{2\pi^2 \hbar^2}
   \left\{ 1 -\frac{k_F ({\boldsymbol r})}{2k_c ({\boldsymbol r})}
   \mbox{ln}\frac{k_c ({\boldsymbol r})+k_F ({\boldsymbol r})}
                 {k_c ({\boldsymbol r})-k_F ({\boldsymbol r})}  \right\} .
\en
Introducing a
{\it running\/} coupling constant implies that the cutoff dependence is
only formal and disappears for sufficiently large
$E_c$~\cite{[Bul02x]}. The cutoff-free superfluid LDA
approach is now in phase of extensive tests~\cite{[Bul02x],[Bor06]}.

\section{Pairing in high-spin isomeric states}

The atomic ground states are strongly correlated. Simple
mean-field-plus-pairing model, although quite successful in
reproducing nuclear masses, see Ref.~\cite{[Lun03]} and
refs. cited therein, cannot incorporate
all important correlations and polarization effects within single
symmetry-broken Slater determinant. Hence,
in order to test/tune less ambiguously basic
theoretical ingredients of our models  is seems natural to
change the attitude and look not into
terribly complicated strongly correlated states but turn the attention
toward as pure as possible physical situations.

The high-spin isomeric states (HSI) open new and
so far not fully
explored possibilities to study, in particular, pair correlations,
blocking phenomena and superfluid-to-normal
phase transition~\cite{[Dra98],[Xu98]}.
They are structurally extremely simple hence both configuration and
shape can be kept rather well under control in the calculations.
In turn, shape and pairing polarization due to blocking
can be studied in detail. As an example let us recall systematic
calculations by Xu {\it et al.\/}~\cite{[Xu99]}
who, using HSI, re-examined traditional average gap
method~\cite{[Mol92]} used to determine
monopole pairing strength $G_{MN}$ for Lipkin-Nogami calculations.
In particular, it was found that inclusion of
polarization effects requires $\sim$10\%
larger pairing strength $G_{MN}$ as compared to the
average-gap value~\cite{[Xu99]} in order
to reproduce experimental data.

\bigskip

Recently the HSI have been systematically observed in $N$=83 nuclei with
60$\leq$Z$\leq$67~\cite{[Oda05]}. In odd-A nuclei the HSI have
$J^\pi$=49/2$^+$ and correspond to a seniority-five, stretched shell-model
configuration [{\bf C}$_o$: $\nu$(f$_{7/2}$h$_{9/2}$i$_{13/2}$)$\otimes$
$\pi$h$_{11/2}$$^2$]$_{49/2}$$^+$.
In odd-odd nuclei
the observed HSI have  $J^\pi$=27$^+$  and are assigned to
[{\bf C}$_{oo}$: $\nu$(f$_{7/2}$h$_{9/2}$i$_{13/2}$)$\otimes$
$\pi$(d$_{5/2}^{-1}$h$_{11/2}$$^2$)]$_{27}$$^+$ configuration.
These two configurations differ by a single
proton hole in Nilsson $[402]5/2$ orbital originating from
spherical $d_{5/2}$ sub-shell. This unique data set
enables to study for the first time OES both at the ground
states (GS) as well as at high-spins using conventional
technique based on binding energy indicators~\cite{[Jen84]}.
The most striking feature of this data set is an almost constant
excitation energy of the HSI what implies that OES at the ground states (GS)
follows very closely the HSI value
$\Delta_{GS}(Z)\approx \Delta_{HSI}(Z)$. Conventional interpretation of
this empirical result in terms of pairing-gap
suggests the lack of blocking phenomenon.
Note, that this
conclusion is independent on the type of pairing indicator.
Hence, in the following, we will use standard three-point
filter $\Delta (Z)$ (see~\cite{[Sat98],[Dob01]} for detailed
discussion of its physical content) of
Eq.~(\ref{3point}).
%\begin{equation}\label{3point}
%  \Delta (Z)=\frac{(-1)^Z}{2}[B(Z-1)+B(Z+1)-2B(Z)].
%\end{equation}

\medskip

\begin{figure}[t]
\par
\centerline{\epsfig{file=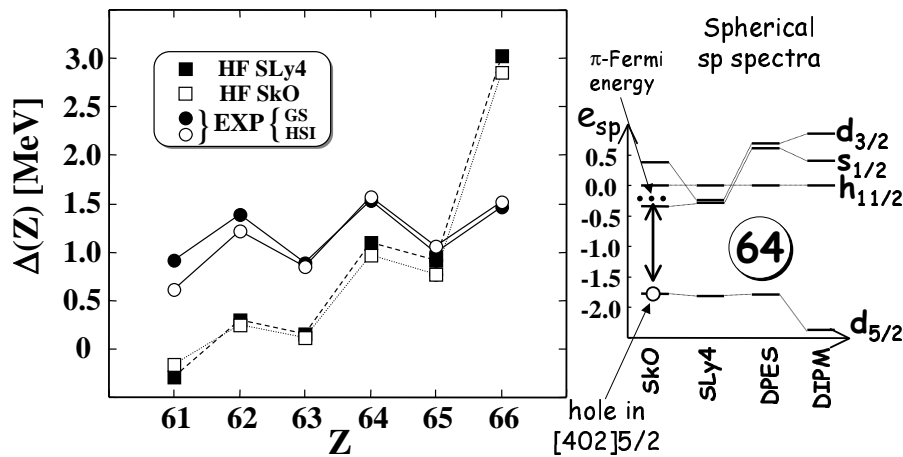,height=6cm,clip}}
\vspace*{8pt}
\caption{Experimental (dots) and theoretical (squares)
OES of Eq.~(\ref{3point}). Calculations for fixed configurations
{\bf C}$_o$ and {\bf C}$_{oo}$ were carried out with single-particle
SHF-SLy4 and SHF-SkO models. Right hand side shows spherical
single-particle spectra of all models discussed in the
context of HSI normalized to $h_{11/2}$.
This part of the figure illustrates schematically a dominant
within the sp scenario
and increasing with $Z$ contribution to OES,
$\Delta(Z) \sim e_ F - e_{[402]5/2}$, originating from
energy difference between "fixed in energy" proton hole $[402]5/2$
and proton Fermi energy which is "moving up" in energy with increasing
$Z$.}
\label{oes_sp}
\end{figure}

The importance of pair-correlations at the HSI can be elucidated by
considering extreme single-particle model.  An example of
such calculations is shown in Fig.~\ref{oes_sp}. The figure compares
experimental OES of Eq.~(\ref{3point}) to theoretical OES computed
using self-consistent SHF model with SLy4 ~\cite{[Cha97]} and
SkO~\cite{[Rei99]} parameterizations. The theoretical values show
clear systematic rise not at all observed in the data.
This theoretical trend is generic and
reflects the fact that, within the extreme sp scenario,
the indicator (\ref{3point}) measures simply a distance between
proton Fermi energy $e_F$ and proton hole state
$e_{[402]5/2}$ as indicated in the right hand side
of the figure. The increase of $\Delta (Z)$ with $Z$ is
therefore generic to any sp model. It can be naturally stopped (or slowed
down) when ph excitations are replaced by quasi-particle
excitations i.e. in the presence of relatively strong pair-correlations.

\begin{figure}[t]
\par
\centerline{\epsfig{file=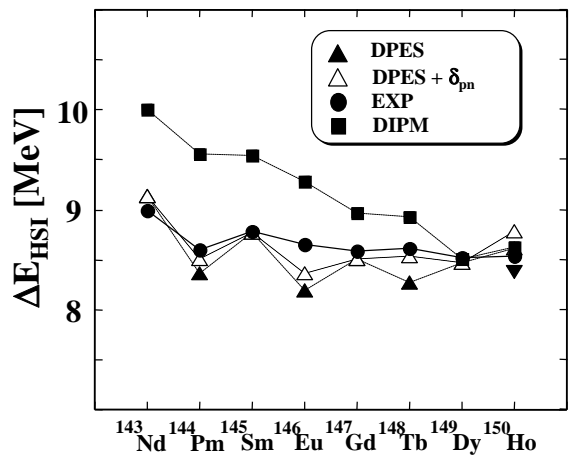,height=6cm,clip}}
\vspace*{8pt}
\caption{Experimental (filled circles) and theoretical (DIPM: squares;
DPES: triangles)
excitation energies of  HSI in the $N$=83 isotones.
Open triangles show the DPES  calculations corrected by
the pn residual interaction $\delta_{pn}$ extracted from nuclear
binding energies  using the 9-point indicator of
Ref.~\protect\cite{[Jen84]}.
}
\label{exc}
\end{figure}

\medskip

To study pairing properties of GS and HSI states we employed
two different approaches: the deformed independent particle model
(DIPM)~\cite{[Dos81]} and the diabatic potential energy surface
model (DPES)~\cite{[Xu98]}.
For the $N$=83 isotones, both  DIPM and  DPES
methods yield  fairly consistent results. Namely, both methods predict:
 ({\it i\/}) weakly
deformed ground states; ({\it ii\/})  well-deformed, oblate ($\beta_2
\approx -0.2$) HSI states
({\it iii\/}) {\bf C$_o$}  and  {\bf
C$_{oo}$} yrast HSI configurations. The only exception is $^{150}$Ho
where DPES predicts  the {\bf C$_{oo}$} configuration to lie
$\sim$150\,keV above the HSI configuration involving four aligned
h$_{11/2}$ protons.
The calculated excitation energies of  HSI, $\Delta E_{HSI}$,
slightly depend on the model used reflecting mostly the
differences in sp spectra of the underlying mean-potentials,  see
Fig.~\ref{oes_sp} and Fig.~\ref{exc}. Note, that
agreement between DPES prediction and experimental $\Delta E_{HSI}$
is excellent particularly after correcting theoretical results for
residual neutron-proton interaction present in the GS of odd-odd nuclei.
Further details concerning both the models and results
can  be found in~\cite{[Oda05]}.

\medskip

\begin{figure}[t]
\par
\centerline{\epsfig{file=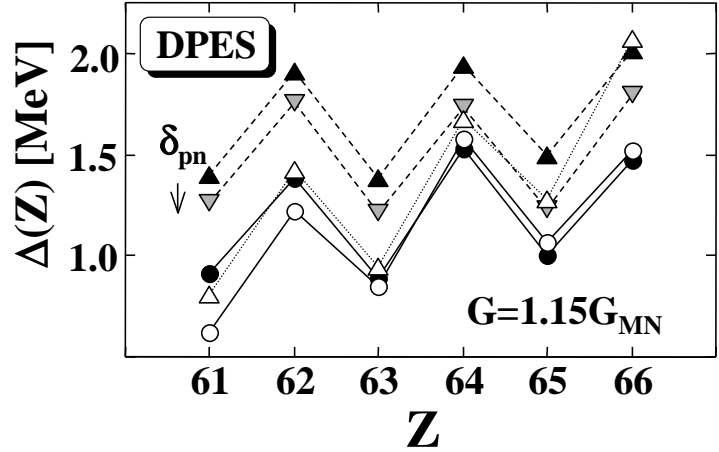,height=6cm,clip}}
\vspace*{8pt}
\caption{$\Delta(Z)$ of  Eq.~(\ref{3point})
calculated using DPES model.
Filled  (open)  symbols denote  GS (HSI) values.
The effect of $\delta_{pn}$ on  GS DPES values is indicated
(gray triangles).
Experimental values of $\Delta(Z)$ are marked by dots.
}
\label{oes}
\end{figure}

The OES of Eq.~(\ref{3point}) calculated using the DPES model
is shown in Fig.~\ref{oes}.
The values calculated at the HSI
follow experimental data very closely meaning that our theoretical
tools and/or interactions can be very reliably tuned out
in such simple situations. On the other hand,
the theory faces serious troubles in reproducing OES in the GS.
The complexity of GS causes that contributions to OES due to
pairing, sp-proton energy splitting,
residual proton-neutron (pn) interaction and/or
nuclear magnetism (time-odd effects) which must all be taken
into account simultaneously in order
to reproduce experimental data are difficult to
control in a satisfactory way in strongly correlated states. In particular,
our calculations clearly indicate that contribution due to blocking which is
mainly responsible for shift of theoretical GS and HSI curves
in Fig.~\ref{oes} is far too
strong and requires revisiting.

\section{Probing isospin dependence of SM matrix elements at
band termination.}

\begin{figure}[t]
\par
\centerline{\epsfig{file=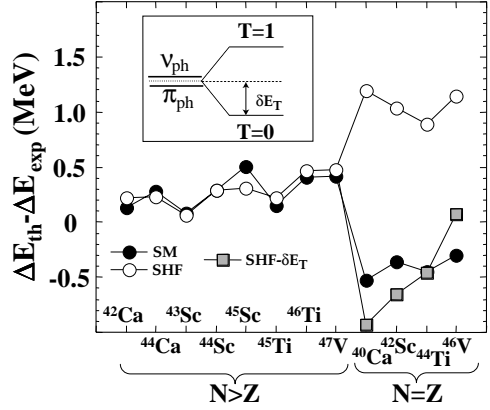,height=6cm,clip}}
\vspace*{8pt}
\caption{Difference  $\Delta E_{\rm th} - \Delta E_{\rm exp}$
between experimental and theoretical values of
$\Delta E$ (\ref{deltae})
in A$\sim$44 mass region.
Dots denote the SM results.
Circles denote the SHF results based on the
modified  SkO
parameterization (see text).
The SHF calculations for the
 $[d_{3/2}^{-1}f_{7/2}^{n+1}]_{I_{max}}$ intruders in
$N$=$Z$ nuclei yield two nearly degenerate states associated with
proton ($\pi_{\rm ph}$) and neutron ($\nu_{\rm ph}$) cross-shell
excitations. As shown in the inset,
the physical $T$=0 state in the laboratory frame
is shifted down  in energy by  $\delta E_T$ (isospin
correlation energy).
Squares denote the SHF results for $N$=$Z$ nuclei
with the isospin correction added.
The SHF results were shifted by 480\,keV in order
to facilitate the comparison with SM.
Taken from Ref.~\protect\cite{[Sto06]}.
}
\label{sko}
\end{figure}

Recently, it was shown \cite{[Zdu05y]} that a
set of terminating states in the mass region A$\sim$ 45 may provide a unique
tool to constrain the Skyrme energy density functional (SEDF). The key idea is
to look into energy differences (\ref{deltae}) between two types of
terminating configurations $[f_{7/2}^n]_{I_{max}}$ and
$[d_{3/2}^{-1} f_{7/2}^{n+1}]_{I_{max}}$
i.e. into the quantity which, as the long experience
in the studies of terminating states within the Nilsson model have clearly
shown, see Ref.~\cite{[Afa99a],[Sat05]} and refs. quoted
therein, belong to the purest
single-particle observables available in nuclear structure. In particular, it
was demonstrated~\cite{[Zdu05y]} that by constraining the
SEDF to the empirical spin-isospin Landau  parameters and by slightly
reducing the  spin-orbit strength, good agreement with  the data could  be
obtained. This result, based on high-spin data for terminating states, is
consistent with conclusions of previous
works~\cite{[Ost92],[Ben02a]} based on
different theoretical methodology and experimental input (such as giant
resonances, beta decays, and moments of inertia). The validation of the
assumption~\cite{[Zdu05y]} regarding the single-particle character of
the maximally-aligned states and the robustness of terminating states in
determining properties of the nuclear EDF was further supported by the recent
comparative study between the fully correlated shell model (SM) and the Skyrme
Hartree-Fock (SHF)~\cite{[Sto06]}, showing essentially a one to one
correspondence between the models at least for $N$$>$$Z$ nuclei, see
Fig.~\ref{sko}.

%------------------------------------------------------------------------------
%

In the $N$=$Z$ nuclei
the comparison between the models can be done only after
evaluation of the correlation energy $\delta E_T$ due to
the  spontaneous breaking of isobaric symmetry in the SHF.
The correction due to isospin symmetry restoration shifts
the $T$=0 state down in energy
in the laboratory system,  as illustrated
in the inset of Fig.~\ref{sko}.
After (phenomenological) symmetry restoration
predictions from both SM and SHF models become very
similar in all considered nuclei. However,
while the agreement in $N$$>$$Z$ is
satisfactory the $N$=$Z$ results deviate strongly from the
data. Definitely too strongly as compared, in particular, to overall
excellent performance of the $fp$ shell SM~\cite{[Cau05]}.

\medskip

One possible origin of this deviation may have its source in
the assumed  truncation to 1p-1h cross-shell excitations.
This configuration-space restriction is expected to  impact the
isoscalar  channel associated with the $sd$$\rightarrow$$fp$ pair
scattering i.e. isoscalar proton-neutron pairing (pn) mode which
is also absent in conventional mean-field description.
A possibility of an onset of the isoscalar pn pairing nearby
band-termination was already discussed within
the mean-field formalism in Ref.~\cite{[Ter98]}.

\medskip

\begin{figure}[t]
\par
\centerline{\epsfig{file=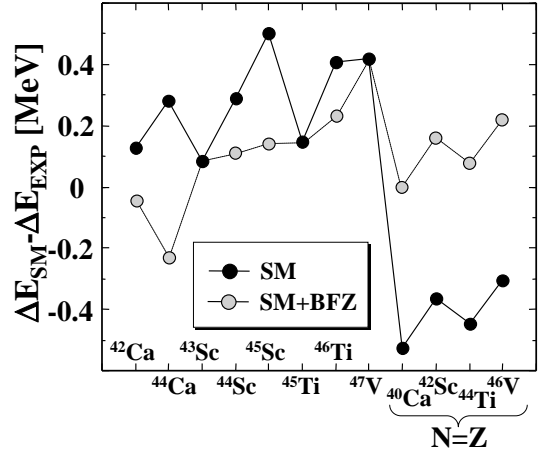,height=6cm,clip}}
\vspace*{8pt}
\caption{
The differences $\Delta E_{SM} - \Delta E_{EXP}$
calculated using SM model (black dots).
Grey dots indicate SM values corrected by
the anticipated contributions due to BFZ effect of Eq.
(\ref{ant-bfz}), assuming $\delta b$$=$700\,keV decrease
in the $b$ strength.
}
\label{fig_bfz}
\end{figure}

An alternative source of deviation may be traced back to incorrect
isospin dependence of SM matrix elements. The quantity (\ref{deltae})
involves particle-hole excitation. In such a case, according to Bansal and
French~\cite{[Ban64]}, and Zamick~\cite{[Zam65]} (BFZ) {\it ... it is a
grave error to assume that ph interaction  is independent on isotopic
spin...}. Under certain assumptions listed in Refs.~\cite{[Ban64],[Zam65]}
the isotopic dependence of ph interaction can be approximated by a
simple
monopole interaction $\sim b {\boldsymbol t}_1 \cdot {\boldsymbol t}_2$
yielding additional contribution to $\Delta E$ of Eq.~(\ref{deltae})
of the form:
\begin{equation}\label{bfz}
\Delta E^{T} = \frac{1}{2} b ( T(T+1) - T_p (T_p+1)  - T_h (T_h+1) ),
\end{equation}
where $T_h=1/2$ and $T_p=T\pm 1/2$ are hole and particle contributions
to the isotopic spin $T$. In the case of
terminating states in $N$=$Z$ nuclei
$T_p= 1/2$ while for terminating states in $N$$>$$Z$ we have $T_p=T - 1/2$.
Hence, the anticipated contributions
to the $\Delta E$ due to the BFZ mechanism are:
\be\label{ant-bfz}
\Delta E^{T}  = \left \{ \ba{lr}  \displaystyle \, -\frac{3}{4}b
& \quad  \textrm{in} \quad N=Z \\    \displaystyle  \quad
            \frac{1}{2}b \left( T-\frac{1}{2} \right) & \quad  \textrm{in}
            \quad N\ne Z \ea
\right.
\ee
If, for some reason,
the strength of the monopole interaction $b$ is overestimated in
the SM by $\delta b$$=$700\,keV  (in this mass region it amounts to
$\delta b/b$$\sim$20\%)
the SM values of $\Delta E$ should be corrected by:  525\,keV in $N$=$Z$
nuclei; 0\,keV in $T=1/2$  $^{43}$Sc, $^{45}$Ti, and $^{47}$V nuclei;
$-$175\,keV in $T=1$ $^{42}$Ca, $^{44}$Sc and $^{46}$Ti nuclei; $-$350\,keV in
$T=3/2$ nucleus $^{45}$Sc; $-$525\,keV in $T=2$ nucleus $^{44}$Ca.
The corrected values are depicted in Fig.~\ref{fig_bfz}.
Note that they match empirical the data almost perfectly.

\medskip

The suggestion that the discrepancy between the SM description of
$N>Z$ and $N=Z$ nuclei may be
caused (fully or partly) by incorrect isospin dependence of the
SM matrix elements~\cite{[Ric91],[War90]} is extremely intriguing and the
anticipated accuracy of $\Delta E_{SM}- \Delta E_{EXP}$ deduced solely on
the basis of the BFZ formula is indeed very
appealing, see Fig.~\ref{fig_bfz}.  To verify the concept we performed
two set of SM calculations using modified SM interactions marked lateron
as SM$^{(1)}$ and  SM$^{(2)}$, respectively.

\begin{figure}[t]
\par
\centerline{\epsfig{file=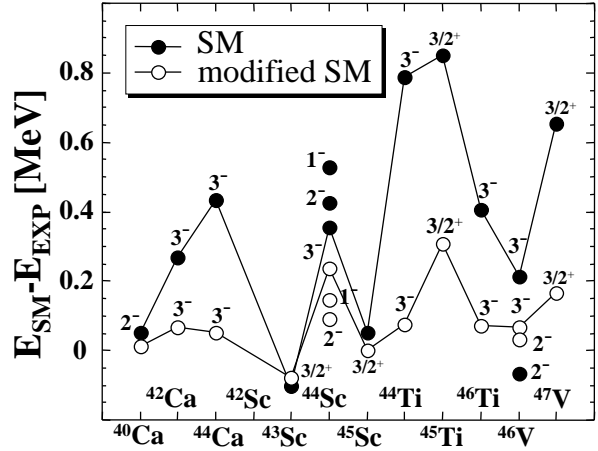,height=6cm,clip}}
\vspace*{8pt}
\caption{Difference between the SM and experimental
excitation energies for selected low-spin 1p-1h intruder states.
Calculations  done
using original SM interaction (bullets) are compared to
the results obtained for modified SM$^{(1)}$ interaction. Note
dramatic improvement obtained for modified SM.
}
\label{intr}
\end{figure}

In the  SM$^{(1)}$ run ALL diagonal particle-hole matrix elements
of original SM interaction~\cite{[Ric91],[War90]} were renormalized
according to the following scheme:
\begin{eqnarray}\label{mod}
& V_{phph} (JT=0)& \rightarrow  V_{phph} (JT=0) +3d  \nonumber \\
& V_{phph} (JT=1)& \rightarrow  V_{phph} (JT=1) - d
\end{eqnarray}
with $d=0.175$keV. The use of
transformation (\ref{mod}) can be partly justified within single $j$-shell
shell-model phenomenological mass formula where it appears to leave invariant
all contributions to the binding energy except the symmetry energy strength
$b_{sym}$ which changes according to the following rule: $b_{sym} \rightarrow
b_{sym} - 4d$, see Ref.~\cite{[Zel96]}. It should be underlined, however, that
$b$ and $b_{sym}$ although having certain common features are different
quantities. Such an attempt to modify $b$ was not fully successful and
satisfactory. On one side we obtained substantial improvement of agreement to
the data for low-lying particle-hole excitations as compared to the original
interaction as shown in Fig.~\ref{intr}.
On the other side the changes appeared not
neutral with respect to other observables spoiling, in particular,
the previously obtained nice agreement
between SM and experimental binding energies.

\smallskip

In the SM$^{(2)}$ run
we have decided to apply renormalization scheme (\ref{mod}) to
CROSS-SHELL matrix elements only. Preliminary
calculations shows that such a simple procedure leads to an almost perfect
agreement between theory and experiment for both terminating as well as
low-lying particle-hole excitations.
Further details concerning our systematic refinements of SM
interaction inspired by very precise high-spin data
including both sets of modifications discussed above will be
published elsewhere~\cite{[Sto07]}.

%------------------------------------------------------------------------------
%
%
%------------------------------------------------------------------------------

\section{Summary and outlook}

It is shown, that local effective theory suitable for low-energy nuclear
structure calculations can be consistently formulated as free form
ultraviolet divergences superfluid local energy density functional
approach. The underlying assumption concerning
independence of infrared physics on ultraviolet dynamics implies
that such a strictly local theory should in principle be
equivalent to finite-range effective theory concerning accuracy
of its predictions. This conclusion makes local realization of
effective theory extremely appealing due to its numerical
simplicity.

\smallskip

There are at least two major problems related to the nuclear LEDF including:
({\it i\/})
Density dependence of LEDF parameters or coupling constants and
its relation to effective
three body NNN interaction. This problem pertains both  to
local as well as to finite-range nuclear EDF formalism in the same way.
({\it ii\/})
Datasets used to parametrize are definitely incomplete leading
to a multitude of local (Skyrme) effective interactions. In this respect
finite-range realizations of effective theories are less prone to
the details of datasets
as our practice with the Gogny interaction shows where
essentially only two parameterizations are used in numerical
applications~\cite{[Dec80x]}. This is related with smaller number of terms
that are used in the expansion (\ref{vcorr}).

\smallskip

The relations between the LEDF parameters and fine-tuning of the LEDF
parameters can be reliably done in simple physical situations.
Long standing experience tells that terminating or isomeric states
belong to the purest known examples of almost unperturbed single-particle
motion~\cite{[Afa99a]}. In this work we summarize our recent efforts
related to terminating and isomeric states
showing, in particular, that:
\begin{itemize}
\item
The structural simplicity of these states can be used
to unify time-odd spin-fields and tune up spin-orbit strength of
the LEDF, see~\cite{[Zdu05y],[Sto06]} for further details.
\item
They can be used for identification,
evaluation and subsequent restoration of broken symmetries inherently
obscuring the SHF treatment, see Ref.~\cite{[Sto06],[Zal06]} for further
details.
\item
Multi quasi-particle
configurations in rare-earth nuclei can be used to test
pairing interaction and blocking phenomena~\cite{[Oda05]}.
\item
Finally, it is shown that these
states appear to be extremely useful in identifying and correcting isobaric
dependence of cross-shell matrix elements
of $sdpf$ SM interaction~\cite{[Sto06],[Sto07]}.
\end{itemize}

\section{Acknowledgments}

The results presented in this manuscript
are due to common effort of many colleagues. I would like to
acknowledge cordially very fruitful collaboration with
D. Dean, M. Kosmulski, W. Nazarewicz, H. Sagawa, G. Stoitcheva, R.A. Wyss,
M. Zalewski, and H. Zdu\'nczuk. This work has been supported
in part by the Polish Committee for Scientific Research (KBN) under
Contract No. 1~P03B~059~27.

%\bibliography{rev}
%\bibliographystyle{unsrt}

\end{document}